\documentclass{article}
\usepackage{graphicx}
\usepackage{amssymb}
\usepackage{amsmath}
\usepackage{color}
\newcommand{\bfr}{\begin{flushright}}
\newcommand{\efr}{\end{flushright}}

\begin{document}

\title{Emergence and violation of geometrical
scaling\\ in pp collisions
\thanks{Presented at  the Low $x$ Workshop, May 30 - June 4 2013, Rehovot and
Eilat, Israel}}
\author{Michal Praszalowicz\footnote{michal@if.uj.edu.pl}
\\
{\small M. Smoluchowski Institute of Physics, Jagellonian University,} \\ 
{\small Reymonta 4,
30-059 Krakow, Poland}
\smallskip\\
}
\date{\today}

\maketitle

\begin{abstract}
We argue that geometrical scaling (GS) proposed originally in the
context of  Deep Inelastic Scattering (DIS) at HERA  works also in pp
collisions at the LHC energies and in NA61/SHINE experiment. We show 
that in DIS  GS
is working up to relatively large Bjorken $x \sim 0.1$. As a consequence 
negative pion multiplicity $p_{\rm T}$
distributions at NA61/SHINE energies exhibit GS in mid rapidity region.
For $y \ne 0$ clear sign of scaling violations
can be seen when one of the colliding partons has Bjorken
$x \ge 0.1$. Finally, we argue that in the case of identified particles GS
scaling is still present but the scaling variable is a function of transverse
mass rather than $p_{\rm T}$. \\
~
\\
PACS number(s):{ 13.85.Ni,12.38.Lg}
\end{abstract}


\section{Introduction}
\label{intro}

This talk based on 
Refs.~\cite{McLerran:2010ex}\nocite{Praszalowicz:2011tc,Praszalowicz:2011rm,
Praszalowicz:2012zh,Praszalowicz:2013uu}--\cite{Praszalowicz:2013fsa} 
(where also an extensive
list of references can be found) follows closely an earlier report of 
Ref.~\cite{Praszalowicz:2013swa}.  
We shall discuss the scaling law, 
called geometrical scaling
(GS), which has been introduced in the context of DIS \cite{Stasto:2000er}. 
It has been also shown that GS is exhibited by the $p_{\text{T}}$
spectra at the LHC \cite{McLerran:2010ex}--\cite{Praszalowicz:2011rm} and that 
an onset of GS can be seen in heavy ion collisions at RHIC energies \cite{Praszalowicz:2011rm}.
At low Bjorken $x<x_{\mathrm{max}}$ gluonic cloud in the proton is  
characterized by an intermediate energy scale $Q_{\text{s}}(x)$,
called saturation scale \cite{sat1,GolecBiernat:1998js}. $Q_{\text{s}}(x)$ is defined as the border
line between dense and dilute gluonic systems  (for review
see \emph{e.g.} Refs.~\cite{Mueller:2001fv,McLerran:2010ub}).
In the present paper we study the consequences of the very existence of $Q_{\text{s}}(x)$;
the details of saturation phenomenon are here not of primary importance.

Here we shall focus of four different pieces of data which exhibit both
emergence and violation of geometrical scaling. In Sect.~\ref{method} we briefly describe
the method used to assess the existence of GS. Secondly, in Sect.~\ref{DIS} we describe
our recent analysis~\cite{Praszalowicz:2012zh} of combined HERA data \cite{HERAdata}
where it has been shown that  GS in DIS 
works surprisingly well up to relatively large $x_{\text{max}}\sim0.1$ 
(see also \cite{Caola:2010cy}). Next, in Sect.~\ref{ppLHC}, on the example of the 
CMS $p_{\rm T}$ 
spectra in central rapidity \cite{Khachatryan:2010xs}, 
we show that  GS is also present in hadronic collisions.  
For particles produced at non-zero rapidities, one
(larger) Bjorken $x=x_{1}$ may be outside of the domain of GS, \emph{i.e.} $x_{1}>x_{\text{max}}$,
and violation of GS should appear. In Sect.~\ref{ppNA61} we present analysis of the
pp data from NA61/SHINE experiment at CERN \cite{NA61} and show that GS is indeed 
violated once rapidity is increased. Finally in Sect. \ref{GSids}
we analyze identified particles spectra where the particle mass provides
another energy scale which may lead to the violation of GS, or at least to some sort of its 
modification \cite{Praszalowicz:2013fsa}.
We conclude in Sect.~\ref{concl}.

\section{Analyzing data with method of ratios}
\label{method}

Geometrical scaling hypothesis means that some observable $\sigma$ depending
in principle on two independent kinematical variables, like $x$ and $Q^2$,
depends in fact only on a given combination of them, denoted in the following as  $\tau$:
\begin{equation}
\sigma(x,Q^{2})=F(\tau)/{Q_{0}^{2}}. \label{GSdef}%
\end{equation}
Here function $F$ in Eq.~(\ref{GSdef}) is a dimensionless universal function of scaling variable
$\tau$:
\begin{equation}
\tau=Q^{2}/Q_{\text{s}}^{2}(x).\label{taudef}%
\end{equation}
and
\begin{equation}
Q_{\text{s}}^{2}(x)=Q_{0}^{2}\left(  {x}/{x_{0}}\right)  ^{-\lambda}
\label{Qsat}%
\end{equation}
is the saturation scale. Here $Q_{0}$ and $x_{0}$ are free parameters which,
however, are not of importance in the present analysis, and exponent 
$\lambda$ is a dynamical quantity of the order of $\lambda\sim0.3$. Throughout
this paper we shall test the hypothesis whether different pieces of data can be described
by formula (\ref{GSdef}) with {\em constant} $\lambda$, and what is the kinematical
range where GS is working satisfactorily. 

As a consequence of Eq.~(\ref{GSdef}) observables $\sigma(x_{i},Q^{2})$ 
for different $x_{i}$'s  should fall on a universal curve, if evaluated  
in terms  of scaling variable  $\tau$. This means 
that  ratios
\begin{equation}
R_{x_{i},x_{\text{ref}}}(\lambda;\tau_{k})=\frac{\sigma%
(x_{i},\tau(x_{i},Q_{k}^{2};\lambda))}{\sigma(x_{\text{ref}%
},\tau(x_{\text{ref}},Q_{k,\text{ref}}^{2};\lambda))} \label{Rxdef}%
\end{equation}  
should be equal to unity independently of $\tau$. Here for some $x_{\rm ref}$
we pick up all $x_i<x_{\rm ref}$ which have at least two overlapping
points in $Q^2$. 

For $\lambda\neq0$ points of the same $Q^{2}$ but
different $x$'s correspond in general to different $\tau$'s. Therefore one has
to interpolate  $\sigma(x_{\text{ref}},\tau(x_{\text{ref}},Q^{2};\lambda))$ to $Q_{k,\text{ref}%
}^{2}$ such that $\tau(x_{\text{ref}},Q_{k,\text{ref}}^{2};\lambda)=\tau_{k}$.
This procedure is described in detail in 
Refs.~\cite{Praszalowicz:2012zh}.

By adjusting $\lambda$ one can make $R_{x_{i},x_{\text{ref}}}(\lambda;\tau
_{k}) \rightarrow 1$ for all $\tau_{k}$ in a given interval.
In order to find an optimal value $\lambda_{\rm min}$ which minimizes
deviations of ratios (\ref{Rxdef}) from unity we form the chi-square measure%
\begin{equation}
\chi_{x_{i},x_{\text{ref}}}^{2}(\lambda)=\frac{1}{N_{x_{i},x_{\text{ref}}}%
-1}{\displaystyle\sum\limits_{k\in x_{i}}}\frac{\left(  R_{x_{i}%
,x_{\text{ref}}}(\lambda;\tau_{k})-1\right)  ^{2}}{\Delta R_{x_{i}%
,x_{\text{ref}}}(\lambda;\tau_{k})^{2}} \label{chix1}%
\end{equation}
where the sum over $k$ extends over all points of given $x_{i}$ that have
overlap with $x_{\text{ref}}$, and ${N_{x_{i},x_{\text{ref}}}}$ is a number of
such points.

\section{Geometrical scaling in DIS at HERA}
\label{DIS}

\begin{figure}[ptb]
\centering
\includegraphics[width=8cm,angle=0]{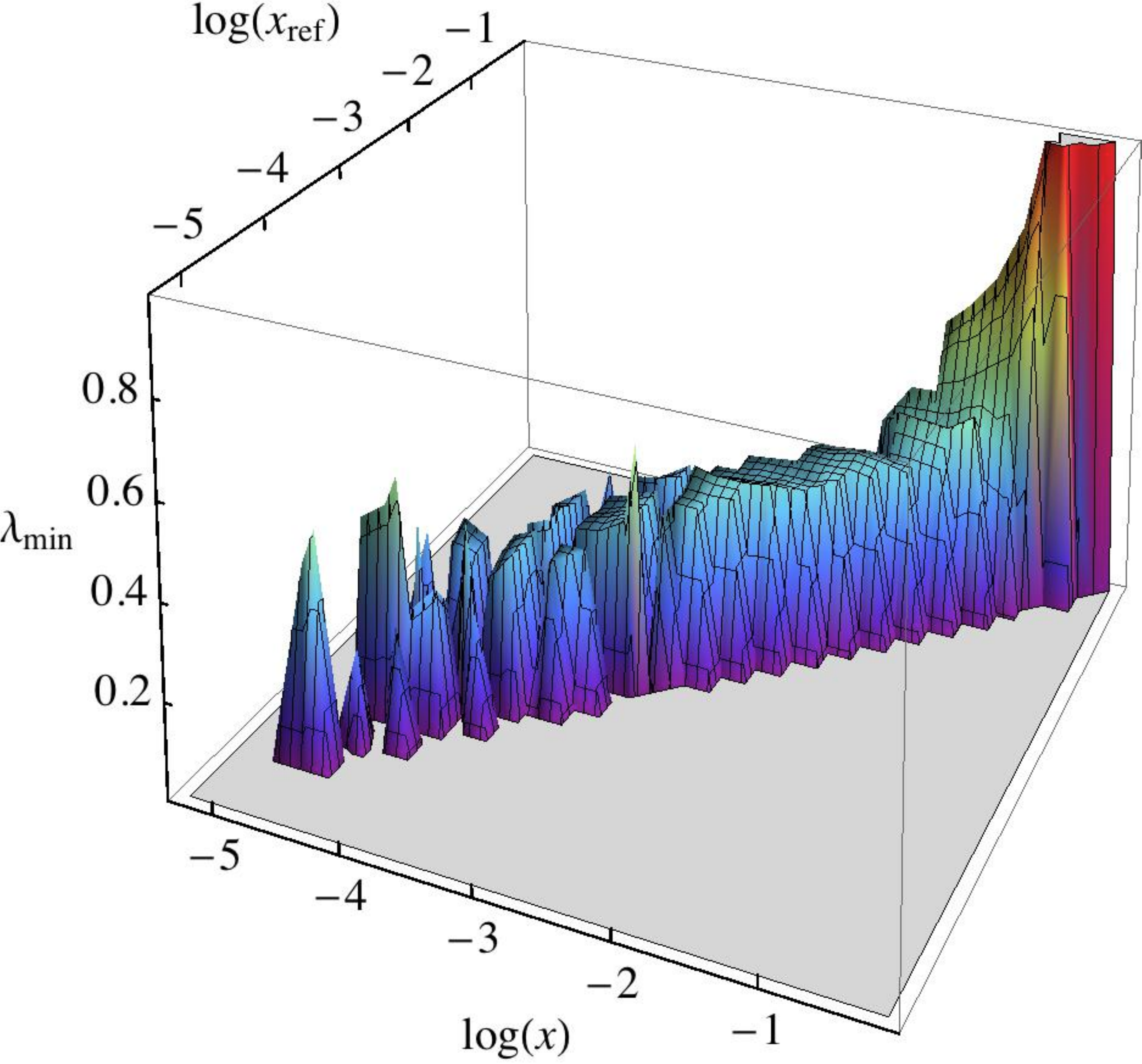}
\caption{Three dimensional
plot of  $\lambda_{\mathrm{min}}(x,x_{\mathrm{ref}})$
obtained by minimization of Eq.~(\ref{chix1}).}%
\label{xlamlog}%
\end{figure}

In the case of DIS the relevant scaling observable is $\gamma^{\ast}p$ cross section
and variable $x$ is simply Bjorken $x$. 
In Fig.~\ref{xlamlog} we
present 3-d plot of $\lambda_{\min}({x,x_{\rm ref}})$ which has been found 
by minimizing  (\ref{chix1}).

Qualitatively,  GS is given by the independence of
$\lambda_{\text{min}}$ on Bjorken $x$ and by the requirement that the
respective  value of $\chi_{x,x_{\text{ref}}}^{2}(\lambda_{\text{min}})$ is
small (for more detailed discussion see Refs.~\cite{Praszalowicz:2012zh}).  
One can see from Fig.~\ref{xlamlog} 
that the stability corner
of $\lambda_{\text{min}}$ 
extends up to $x_{\text{ref}}\lesssim0.1$,
which is well above the original expectations.
In Ref.~\cite{Praszalowicz:2012zh} we have shown that:
\begin{equation}
\lambda = 0.32 - 0.34 \,\,\,\,\, {\rm for} \,\,\,\,\, x \le 0.08.
\end{equation}

\section{Geometrical scaling of central rapidity $p_{\rm T}$ spectra at the LHC}  
\label{ppLHC}

\begin{figure}[h]
\centering
\includegraphics[scale=0.70]{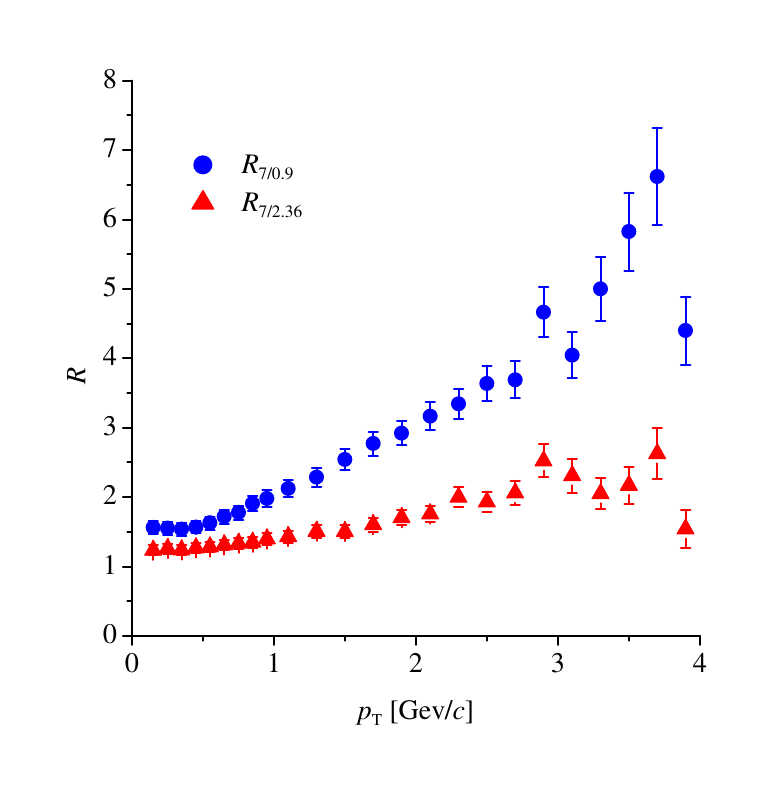} %
\includegraphics[scale=0.70]{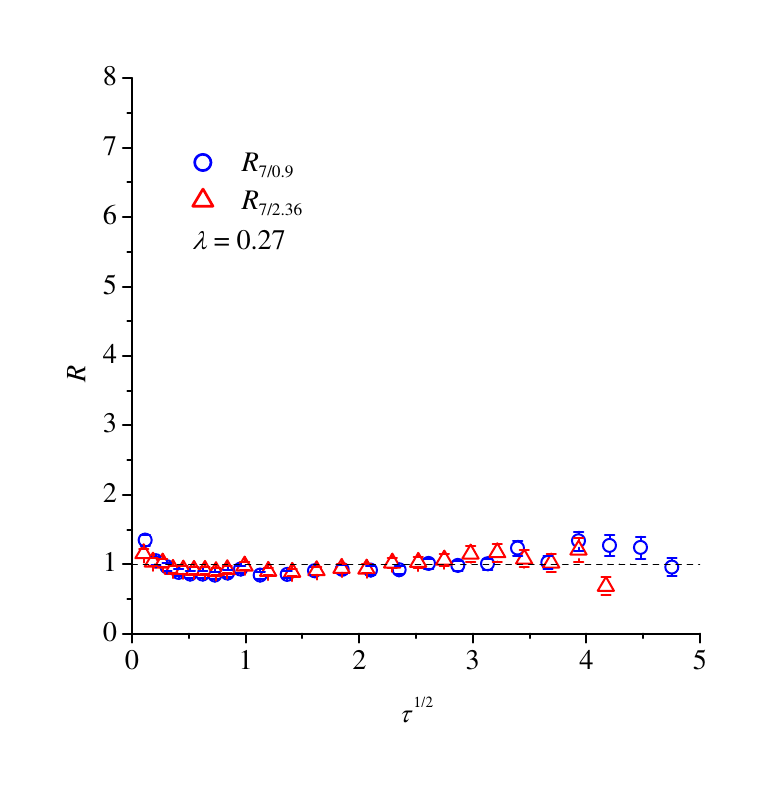}
\caption{Ratios of CMS $p_{\mathrm{T}}$ spectra \protect\cite%
{Khachatryan:2010xs} at 7 TeV to 0.9 (blue circles) and 2.36 TeV (red
triangles) plotted as functions of $p_{\mathrm{T}} $ (left) and scaling
variable $\protect\sqrt{\protect\tau}$ (right) for $\protect\lambda=0.27$. }
\label{ratios1}
\end{figure}

In hadronic collisions at c.m. energy $W=\sqrt{s}$ particles are produced in 
the scattering process of
two patrons (mainly gluons) carrying Bjorken $x$'s
\begin{equation}
x_{1,2}=e^{\pm y}\,p_{\text{T}}/W.\label{x12}%
\end{equation}
For central rapidities $x=x_1 \sim x_2$. In this case 
charged particles multiplicity spectra exhibit GS
\cite{McLerran:2010ex}
\begin{equation}
\left.  \frac{dN}{dy d^{2}p_{\text{T}}}\right\vert _{y\simeq0}=\frac
{1}{Q_{0}^{2}}F(\tau)\label{GSinpp}%
\end{equation}
where $F$ is a universal dimensionless function of the scaling variable (\ref{taudef}).
Therefore the method of ratios can be applied to the multiplicity distributions at different
energies ($W_i$ taking over the role of $x_i$ in Eq.~(\ref{Rxdef}))\footnote{
For pp collisions we define ratios $R_{W_{\rm ref},W_i}$ as an inverse of (\ref{Rxdef})}. For
$W_{\rm ref}$ we take the highest LHC energy of 7~TeV. Hence one can
form two ratios $R_{W_{\rm ref},W_i}$ with $W_1 =2.36$ and $W_2 = 0.9$~TeV.
These ratios are plotted in Fig.~\ref{ratios1} for the CMS single non-diffractive spectra
for $\lambda = 0$ and for $\lambda = 0.27$, which minimizes (\ref{chix1})
in this case. We see that original ratios plotted in terms of $p_{\text{T}%
}$ range from 1.5 to 7, whereas plotted in terms of $\sqrt{\tau}$ they are
well concentrated around unity. The optimal exponent $\lambda$ is, however,
smaller than in the case of DIS. Why this so, remains to be understood.

\section{Violation of geometrical scaling in forward rapidity region}
\label{ppNA61}

For $y >0$ two Bjorken $x$'s can be quite different:
$x_{1}>x_{2}$. Therefore by increasing $y$ one can
eventually reach $x_{1}>x_{\mathrm{max}}$ 
and GS violation should be seen.
For that purpose we shall use
pp data from NA61/SHINE experiment at CERN \cite{NA61}
at different rapidities $y=0.1-3.5$ and at five scattering
energies $W_{1,\ldots,5}=17.28,\;12.36,\;8.77,\;7.75$, and $6.28$ GeV.

\begin{figure}[h!]
\centering
\includegraphics[width=6.0cm,angle=0]{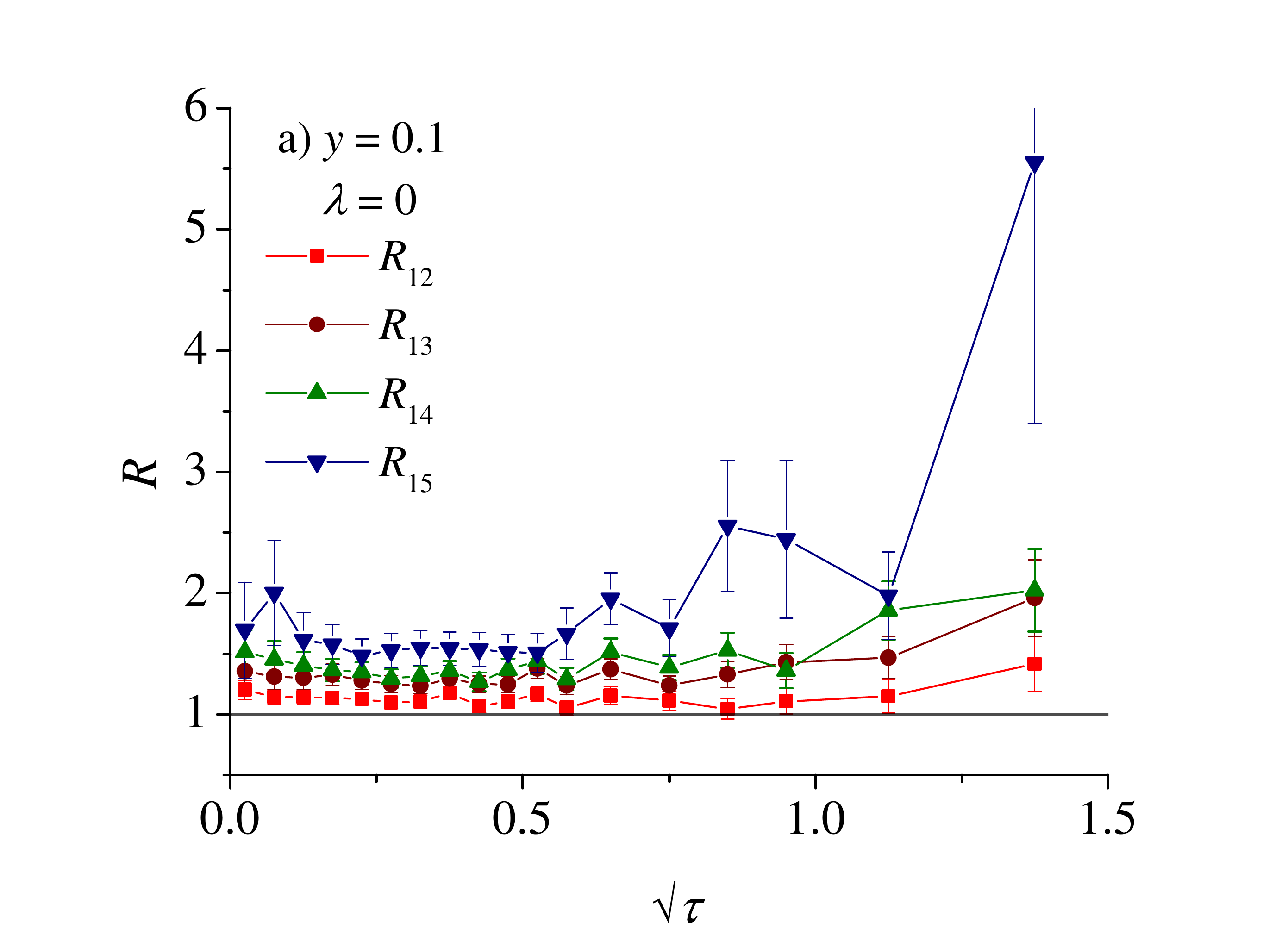}
\includegraphics[width=6.0cm,angle=0]{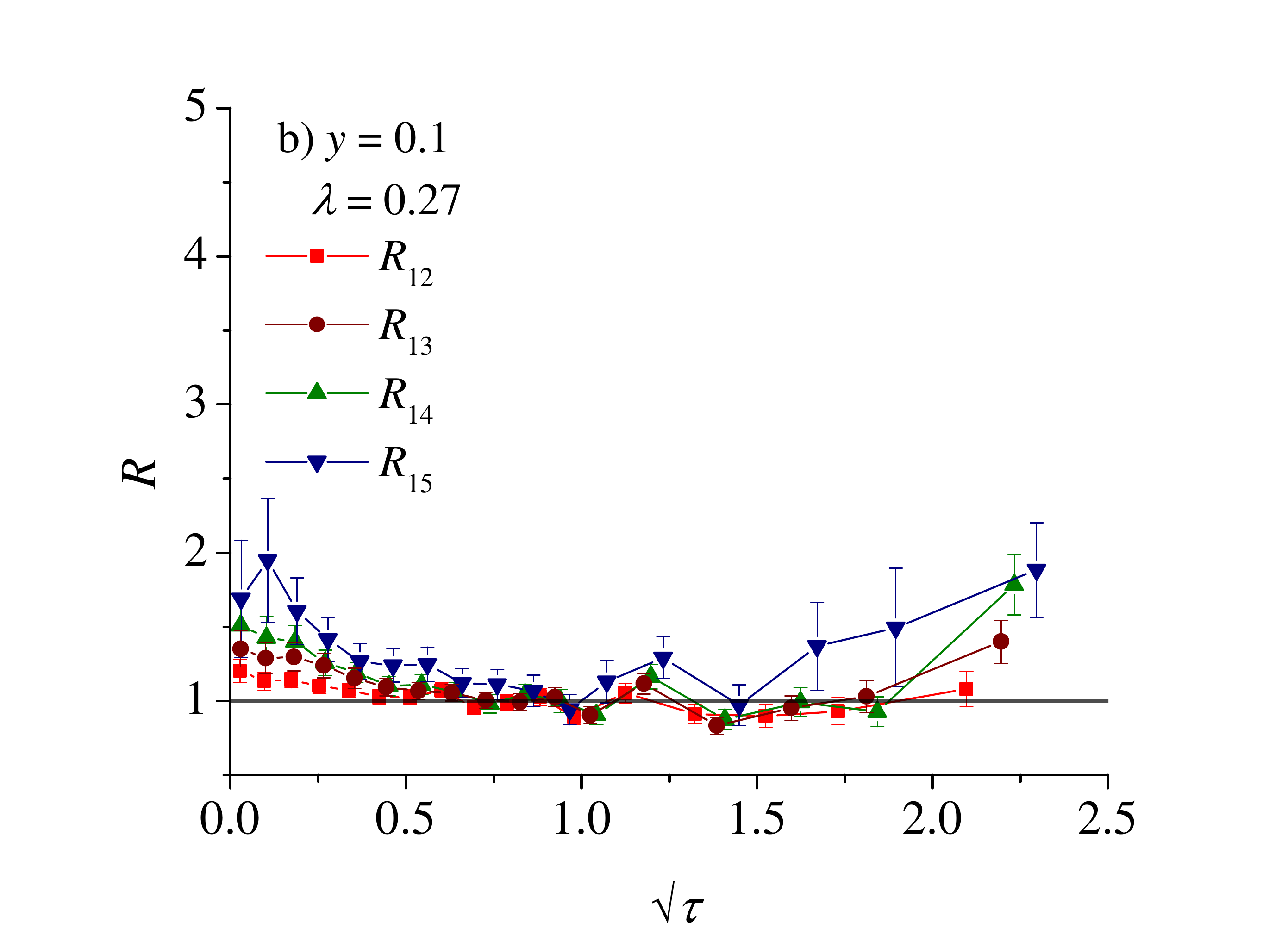} 
\caption{Ratios $R_{1k}$
as functions of $\sqrt{\tau}$ for the lowest rapidity $y=0.1$: a) for
$\lambda=0$ when $\sqrt{\tau}=p_{\mathrm{T}}$ and b) for $\lambda=0.27$ which
corresponds to GS.}%
\label{y01}%
\end{figure}

\begin{figure}[h!]
\centering
\includegraphics[width=6.0cm,angle=0]{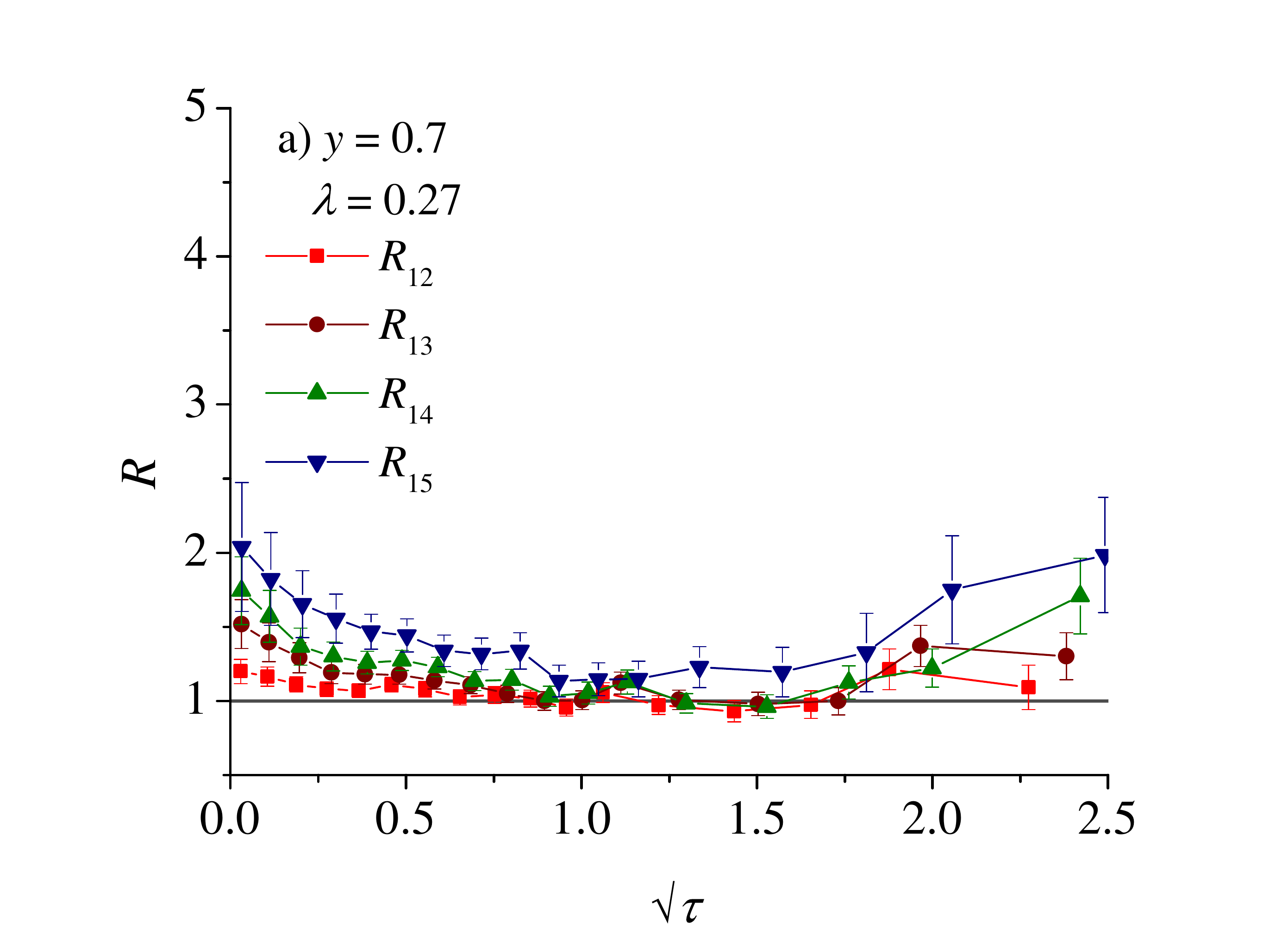}
\includegraphics[width=6.0cm,angle=0]{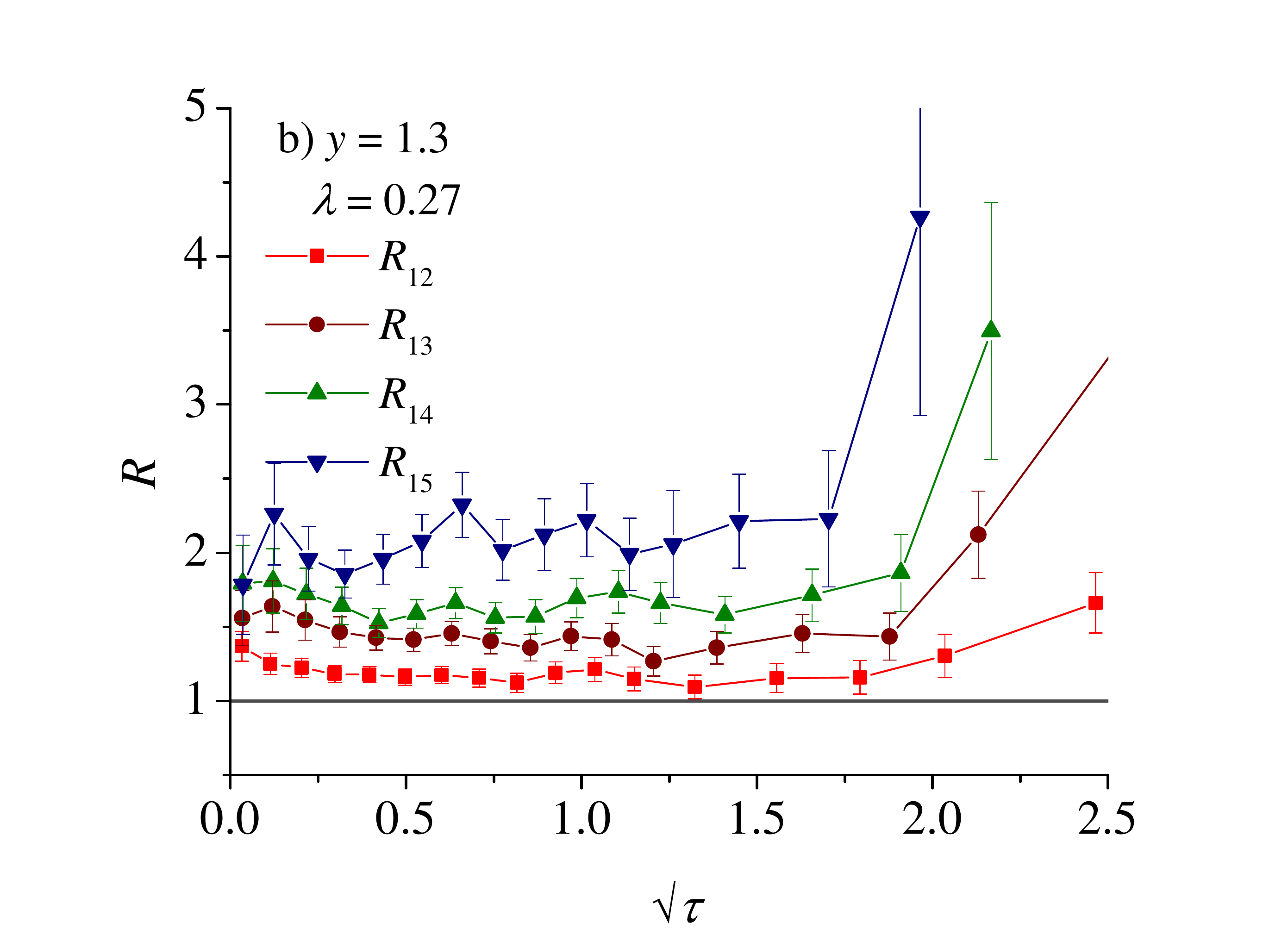} 
\caption{Ratios $R_{1k}$
as functions of $\sqrt{\tau}$ for $\lambda=0.27$ and for different rapidities
 a) $y=0.7$ and b) $y=1.3$. With increase of rapidity,
gradual closure of the GS window can be seen.}%
\label{ys}%
\end{figure}

In Fig.~\ref{y01} we plot ratios $R_{1i}=R_{W_1,W_i}$ (\ref{Rxdef}) for $\pi^-$ spectra in
central rapidity 
for $\lambda=0$ and 0.27. 
For $y=0.1$  the GS region extends down to the smallest energy
because $x_{\rm max}$ is as large as 0.08.
However, the quality of GS is the worst for the lowest energy $W_5$.
 By increasing $y$  some points fall outside  the GS window because  $x_1 \ge x_{\rm max}$, 
 and finally for
$y\ge1.7$ no GS is present in NA61/SHINE data. This is illustrated 
nicely in Fig.~\ref{ys}.

\section{Geometrical scaling for identified particles}
\label{GSids}

In Ref.~\cite{Praszalowicz:2013fsa} we have proposed that in the case of 
identified particles another scaling
variable should be used in
which $p_{\mathrm{T}}$ is replaced by 
$\tilde{m}_{\mathrm{T}} =m_{\rm T} - m =  \sqrt{m^2_{\rm T} +p^2_{\rm T}}-m $ 
($\tilde
{m}_{\mathrm{T}}$ -- scaling), \emph{i.e.}
\begin{equation}
\tau_{\tilde{m}_{\rm T}} =\frac{\tilde{m}_{\text{T}}^{2}}{Q_{0}^{2}}\left(
\frac{\tilde{m}_{\text{T}}}{x_0 W}\right)  ^{\lambda}.\label{taumtdef}%
\end{equation}
This choice is purely phenomenological for the following reasons. Firstly, the
gluon cloud is in principle not sensitive to the mass of the particle it
finally is fragmenting to, so in principle one should take $p_{\mathrm{T}}$ as
an argument of the saturation scale. In this case the proper scaling variable
would be
\begin{equation}
\tau_{\tilde{m}_{\rm T} p_{\rm T}} =\frac{\tilde{m}_{\text{T}}^{2}}{Q_{0}^{2}}\left(
\frac{p_{\text{T}}}{x_0 W}\right)  ^{\lambda}.\label{taumtpdef}%
\end{equation}
However this choice ($\tilde{m}_{\mathrm{T}}%
$$p_{\mathrm{T}}$ -- scaling) does not really differ numerically from the one
given by Eq.~(\ref{taumtdef}). 

\begin{figure}[h!]
\centering
\includegraphics[scale=0.3]{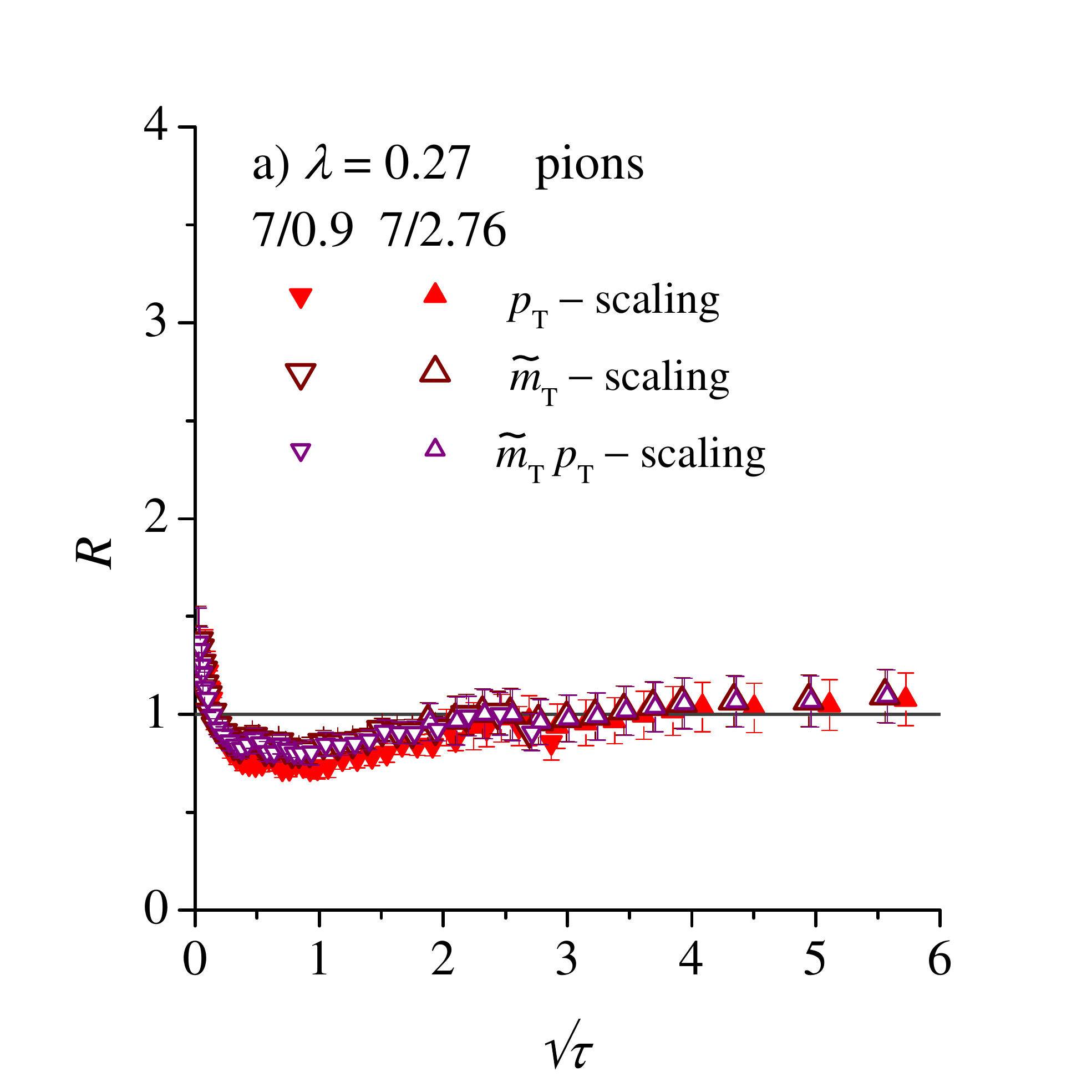}
\includegraphics[scale=0.3]{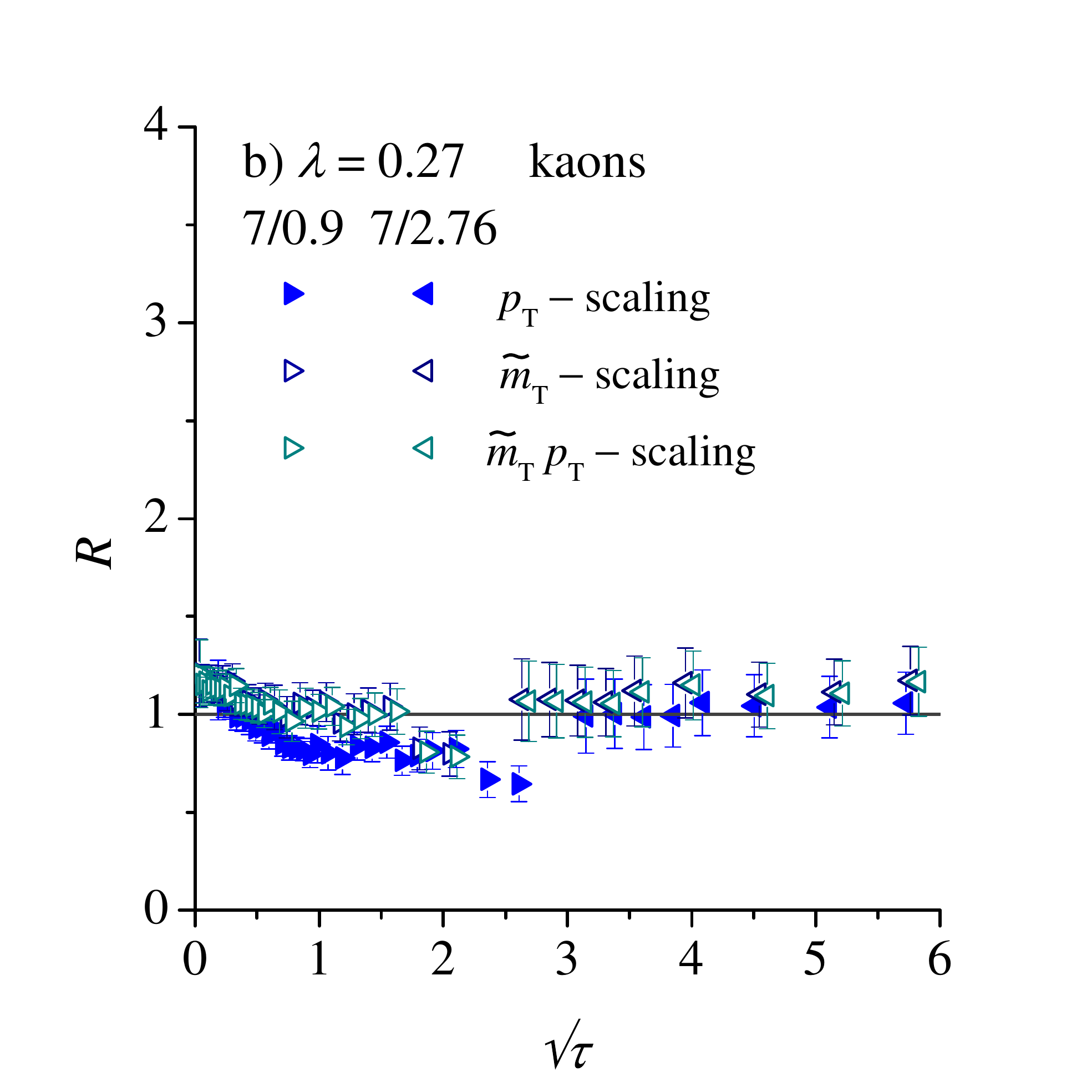}\\
\includegraphics[scale=0.3]{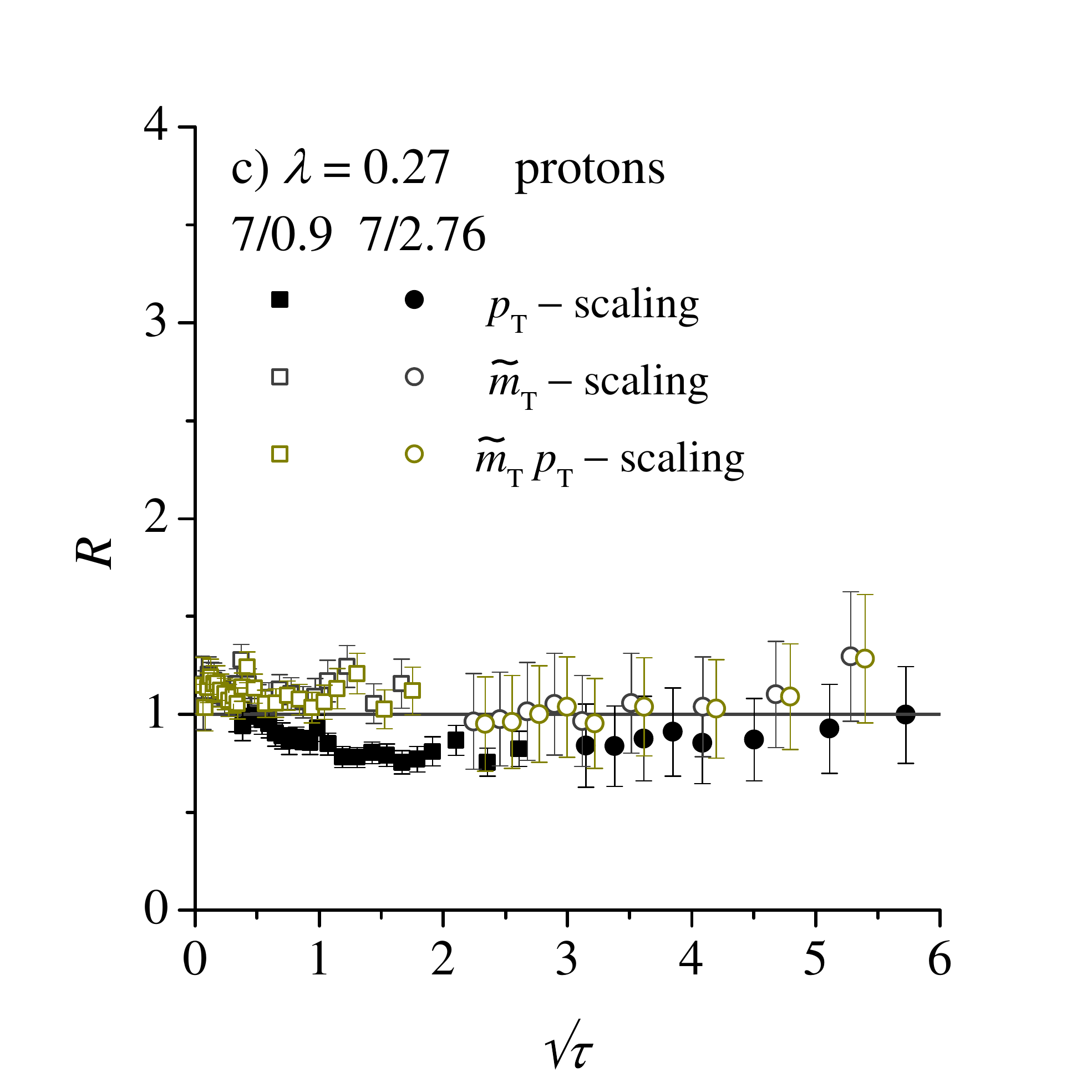}
\includegraphics[scale=0.3]{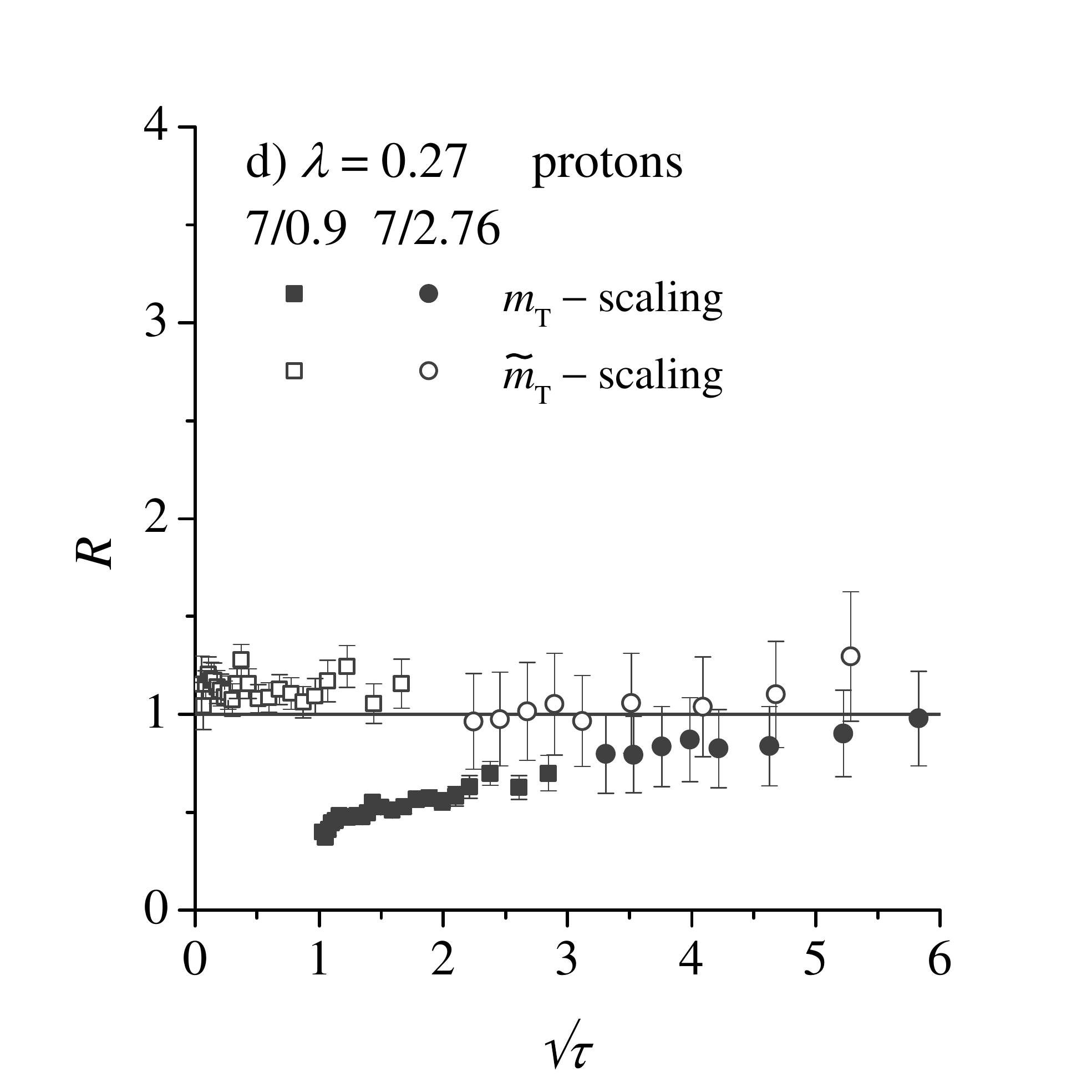}
\caption{Panels a) -- c):
comparison of geometrical scaling in three different variables: $\tau_{p_{\rm T}}$,
$\tau_{\tilde{m}_{\rm T}}$ and $\tau_{\tilde{m}_{\rm T} p_{\rm T}}$ 
for $\lambda=0.27$. Full symbols
correspond to ratios $R_{W_{1}/W_{2}}$ plotted in terms of the scaling
variable $\tau_{p_{\rm T}}$, open symbols to $\tau_{\tilde{m}_{\rm T}}$ 
and $\tau_{\tilde{m}_{\rm T}%
p_{\rm T}}$, note negligible differences between the latter two forms of scaling
variable. Panel a) corresponds to pions, b) to kaons and c) to protons. In
panel d) we show comparison of geometrical scaling for protons in scaling
variables $\tau_{\tilde{m}_{\rm T}}$ and $\tau_{m_{\rm T}}$, 
no GS can be achieved in the
latter case.}%
\label{ratiosmandpT}%
\end{figure}

To this end let us see how scaling properties of GS are affected by
going from scaling variable $\tau_{p_{\rm T}}=\tau$ (\ref{taudef}) to $\tau_{\tilde
{m}_{\mathrm{T}}}$ (\ref{taumtdef}) and what would be the difference in
scaling properties if we had chosen $p_{\mathrm{T}}$ as an argument in the
saturation scale leading to scaling variable $\tau_{\tilde{m}_{\mathrm{T}%
}p_{\mathrm{T}}}$ (\ref{taumtpdef}). This is illustrated in Fig.~\ref{ratiosmandpT}
where we show analysis \cite{Praszalowicz:2013fsa} of recent ALICE data 
on identified particles \cite{ALICE}. In Fig.~\ref{ratiosmandpT}.a-c
full symbols refer to the $p_{\mathrm{T}}$ --
scaling (\ref{taudef}) and open symbols to $\tilde{m}_{\mathrm{T}}$ --
scaling or $\tilde{m}_{\mathrm{T}}$$p_{\mathrm{T}}$ -- scaling. One can see
very small difference between open symbols indicating that scaling variables
$\tau_{\tilde{m}_{\mathrm{T}}}$ (\ref{taumtdef}) and $\tau_{\tilde
{m}_{\mathrm{T}}p_{\mathrm{T}}}$ (\ref{taumtpdef}) exhibit GS of the same
quality. On the contrary $p_{\mathrm{T}}$ -- scaling in variable $\tau_{p_{\rm T}}$
(\ref{taudef}) is visibly worse.

Finally in Fig.~\ref{ratiosmandpT}.d, on the example of protons, we compare
$\tilde{m}_{\mathrm{T}}$ -- scaling (open symbols) and ${m}_{\mathrm{T}}$ -- scaling 
(full symbols) in variable
\begin{equation}
\tau_{m} =\frac{{m}_{\text{T}}^{2}}{Q_{0}^{2}}\left(  \frac{m_{\text{T}}}%
{x_0 W}\right)  ^{\lambda}.\label{taumdef}%
\end{equation}
for
$\lambda=0.27$. One can see that no GS has been achieved in the latter case.
Qualitatively the same behavior can be observed for other values of $\lambda$.

\section{Conclusions}
\label{concl}

In Ref.~\cite{Praszalowicz:2012zh} we  have shown that GS in DIS works well 
up to rather large Bjorken $x$'s
with exponent $\lambda = 0.32 - 0.34$.
In pp collisions at the LHC energies in central rapidity GS is seen in 
the charged particle multiplicity spectra, however, $\lambda = 0.27$ in this case
\cite{McLerran:2010ex}. 
By changing rapidity  one can force one of the Bjorken $x$'s of colliding patrons
to exceed $x_{\rm max}$ and GS violation is expected. Such behavior is indeed
observed in the NA61/SHINE pp data \cite{Praszalowicz:2013uu}. Finally
we have shown that for identified particles scaling variable $\tau$ of Eq.~(\ref{taudef})
should be replaced by $\tau_{\tilde{m}_{\rm T}}$ defined in Eq.~(\ref{taumtdef}) and
the scaling exponent $\lambda \approx 0.3$ 
\cite{Praszalowicz:2013fsa}.

\bigskip
\section*{Acknowledgemens}
Many thanks 
are due to the organizers of this successful series of conferences.
This work was supported by 
the Polish NCN  grant 2011/01/B/ST2/00492.


\bigskip

\begin{thebibliography}{99}
\bibitem{McLerran:2010ex} 
  L.~McLerran and M.~Praszalowicz,
  Acta Phys.\ Pol.\ B {\bf 41} (2010) 1917
  and
  Acta Phys.\ Pol.\ B {\bf 42} (2011) 99.
  
\bibitem{Praszalowicz:2011tc} 
  M.~Praszalowicz,
  Phys.\ Rev.\ Lett.\  {\bf 106} (2011) 142002.
\bibitem{Praszalowicz:2011rm} 
  M.~Praszalowicz,
  Acta Phys.\ Pol.\ B {\bf 42} (2011) 1557
  and
arXiv:1205.4538 [hep-ph].
  

\bibitem{Praszalowicz:2012zh} 
  M.~Praszalowicz  and T.~Stebel,
  JHEP {\bf 1303} (2013) 090  and {\bf 1304} (2013) 169.
  
  \bibitem{Praszalowicz:2013uu} 
  M.~Praszalowicz,
  Phys. Rev. D {\bf87} (2013) 071502(R).
 
\bibitem{Praszalowicz:2013fsa}
  M.~Praszalowicz,
  arXiv:1308.5911 [hep-ph].
  
\bibitem{Praszalowicz:2013swa}
  M.~Praszalowicz,
  Acta Phys. Pol. B Proceeding Supplement {\bf 6} (2013) 815.
  
\bibitem{Stasto:2000er} 
  A.~M.~Stasto, K.~J.~Golec-Biernat and J.~Kwiecinski,
  Phys.\ Rev.\ Lett.\  {\bf 86} (2001) 596.
  

\bibitem{sat1}
L. V. Gribov, E. M. Levin and M. G. Ryskin, Phys. Rept. {\bf 100} (1983) 1; \\
A. H. Mueller and  J-W. Qiu, Nucl. Phys. {\bf 268} (1986) 427; 
A. H. Mueller, Nucl. Phys. {\bf B558} (1999) 285.   

\bibitem{GolecBiernat:1998js} 
  K.~J.~Golec-Biernat and M.~W{\"u}sthoff,
  Phys.\ Rev.\ D {\bf 59} (1998) 014017
  and
  Phys.\ Rev.\ D {\bf 60} (1999) 114023.
  
  \bibitem {Mueller:2001fv}A.~H.~Mueller, \emph{Parton Saturation: An Overview},
arXiv:hep-ph/0111244.

\bibitem {McLerran:2010ub}L.~McLerran, 
{Acta Phys.\ Pol.\ B} \textbf{41} (2010) 2799.  

\bibitem {HERAdata}C.~Adloff \textit{et al.} [H1 Collaboration],
{Eur.\ Phys.\ J.\ C} \textbf{21} (2001) 33; 
S.~Chekanov \textit{et al.} [ZEUS Collaboration],
{Eur.\ Phys.\ J.\ C} \textbf{21} (2001) 443. 

\bibitem{Caola:2010cy} 
  F.~Caola, S.~Forte and J.~Rojo,
  Nucl.\ Phys.\ A {\bf 854} (2011) 32.
  

\bibitem{Khachatryan:2010xs}  V.~Khachatryan \textit{et al.} [CMS
Collaboration], 
JHEP \textbf{1002} (2010) 041 
and 
Phys.\ Rev.\ Lett.\ \textbf{105} (2010) 022002
and 
JHEP \textbf{1101} (2011) 079. 
%


  \bibitem{NA61}
 N. Abgrall {\em et al.} [NA61/SHINE Collaboration],
 {\em Report from the NA61/SHINE experiment
at the CERN SPS} CERN-SPSC-2012-029, SPSC-SR-107;\\ 
 A. Aduszkiewicz, Ph.D. Thesis in prepartation, University of Warsaw, 2013;\\
 Sz. Pulawski, talk at 9th Polish Workshop on Relativistic Heavy-Ion Collisions,
Krak{\'o}w, November 2012
 and private communication.
 
 \bibitem {ALICE} K.~Aamodt \textit{et al.} [ALICE Collaboration],
Eur.\ Phys.\ J.\ C \textbf{71} (2011) 1655  [arXiv:1101.4110 [hep-ex]];
\newline A.~Ortiz Velasquez [ALICE Collaboration],
Nucl.\ Phys.\  A \textbf{904-905} (2013) 763c  [arXiv:1210.6995 [hep-ex]]
(ALICE preliminary).

 
\end{thebibliography}
\end{document}